# Magnetoelectric Coupling by Giant Piezoelectric Tensor Design

J. Irwin[a,1], S. Lindemann[b,1], W. Maeng[b], J. J. Wang[c], V. Vaithyanathan[d], J.M. Hu[b],
L.Q. Chen[c], D.G. Schlom[d,e], C.B. Eom[b], M.S. Rzchowski[a,2]

[a]Department of Physics, University of Wisconsin-Madison
Madison, Wisconsin 53706, United States

[b]Department of Materials Science and Engineering, University of Wisconsin-Madison
Madison, Wisconsin 53706, United States

[c]Department of Materials Science and Engineering, Pennsylvania State University, University
Park, Pennsylvania 16802, United States

[d]Department of Material Science and Engineering, Cornell University,
Ithaca, New York 14853, United States

[e]Kavli Institute at Cornell for Nanoscale Science,
Ithaca, New York 14853, United States

## Abstract

Strain-coupled magnetoelectric (ME) phenomena in piezoelectric / ferromagnetic thin-film bilayers are a promising paradigm for sensors and information storage devices, where strain is utilized to manipulate the magnetization of the ferromagnetic film. In-plane magnetization rotation with an electric field across the film thickness has been challenging due to the virtual elimination of in-plane piezoelectric strain by substrate clamping, and to the requirement of anisotropic in-plane strain in two-terminal devices. We have overcome both of these limitations by fabricating lithographically patterned devices with a piezoelectric membrane on a soft substrate platform, in which in-plane strain is freely generated, and a patterned edge constraint that transforms the nominally isotropic piezoelectric strain into the required uniaxial strain. We fabricated 500 nm thick, (001) oriented $[Pb(Mg_{1/3}Nb_{2/3})O_3]_{0.7}$-$[PbTiO_3]_{0.3}$ (PMN-PT) unclamped piezoelectric membranes with ferromagnetic Ni overlayers. Guided by analytical and numerical continuum elastic calculations, we designed and fabricated two-terminal devices exhibiting Ni magnetization rotation in response to an electric field across the PMN-PT. Similar membrane heterostructures could be used to apply designed strain patterns to many other materials systems to control properties such as superconductivity, band topology, conductivity, and optical response.

**Magnetoelectric coupling | piezoelectric | strain | membrane**

[1]J.I. and S.L. contributed equally to this work
[2]To whom correspondence should be addressed. E-mail: rzchowski@physics.wisc.edu

**Introduction**

Magnetoelectric materials systems possess a wide range of applications including non-volatile memories, magnetic field sensors, spintronics, tunable RF circuit elements, tunable optics, and biomedical devices (1–3). Significant effort has been devoted towards the few known materials exhibiting single-phase room temperature magnetoelectricity, but these materials have drawbacks such as weak magnetoelectric coupling or small electric polarizations (4). Composite magnetoelectrics, consisting of a ferromagnet coupled to a piezoelectric via strain, are a well-studied alternative to single phase magnetoelectrics. Composite magnetoelectrics have the largest reported magnetoelectric coupling constants and suitable electric polarizations, magnetic coercive fields, and saturation magnetizations. These characteristics make them highly promising device candidates, but up to this point they have been challenging to implement in thin-film form (2, 5–7). In this work we design, fabricate, and characterize (001)-oriented, epitaxial thin-film magnetoelectric membrane heterostructures based on the piezoelectric material $[Pb(Mg_{1/3}Nb_{2/3})O_3]_{0.7}$-$[PbTiO_3]_{0.3}$ (PMN-PT) (8, 9). Giant piezoelectric coefficients and large electro-mechanical coupling have allowed PMN-PT based composite magnetoelectrics to achieve superior performance, for example in magnetic field sensors (3). By designing the piezoelectric tensor we overcome previous limitations intrinsic to thin films, and demonstrate electric field control of in-plane magnetization at low bias voltages.

The structure presented here overcomes two critical factors limiting thin-film composite magnetoelectrics. The first limitation arises from substrate clamping that virtually eliminates the in-plane piezoelectric response of thin films, and the second limitation arises from the in-plane four-fold symmetry of most (001) piezoelectrics that precludes the anisotropic in-plane strain necessary for in-plane magnetization rotation. Substrate clamping has limited the majority of composite magnetoelectric research to bulk piezoelectrics (7, 10). Nanoscale patterning has been shown to partially address this by relaxing the island through its thickness (11), but such nano-structuring can introduce unwanted defects such as ion implantation and dislocations (12). Membranes are free from substrate clamping and operate at low voltage while still providing for high device density. Special crystalline orientations (13, 14), domain switching (15–17), and extra top electrodes (18) have addressed the in-plane symmetry limitation, but our piezoelectric tensor design approach directly transforms biaxial piezoelectric strain into uniaxial strain that reorients in-plane magnetization, eliminating complexity and fabrication challenges. We demonstrate in-plane magnetization reorientation with out-of-plane electric fields, and develop design principles that can be used to generate specific strain patterns.

**Experimental Approach**

The membrane fabrication process starts from an epitaxial PMN-PT / SrRuO$_3$ bilayer on SrTiO$_3$-buffered Si, and results in a piezoelectric membrane heterostructure on a soft polymer (Polydimethylsiloxane [PDMS]) coated glass slide (Fig. 1, see Materials and Methods for details). Growth of high quality epitaxial PMN-PT/SrRuO$_3$/SrTiO3 heterostructures on 4° miscut (001)-oriented Si substrates has been previously reported (19). A continuous Pt film sputtered onto the PMN-PT serves as the bottom electrode. The structure is attached Pt side down to soft PDMS coated glass, then the Si substrate is removed with a XeF$_2$ plasma etch, and the SiO$_2$ by ion-milling. This leaves behind a sub-micron thick PMN-PT / SrRuO$_3$. The exposed SrRuO$_3$ is patterned into top electrodes, defining the PMN-PT biased regions. A 35 nm thick Ni layer is deposited and patterned into regions in which we probe the magnetization rotation via Magneto-optic Kerr Effect (MOKE) measurements. A protective coating of SU-8 polymer and an overlayer of patterned Au allows probe tips to contact individual top electrodes. The cross section of the final heterostructure is shown in Fig. 2A.

Structural, ferroelectric and piezoelectric characterization of the PMN-PT was performed on thin-film and membrane samples. High-resolution four-circle X-ray diffraction shows that the biaxially strained thin film PMN-PT relaxes towards bulk lattice constants after substrate removal (SI Appendix, Fig. S1B). Release from the substrate also results in a slight increase in the PMN-PT (002) rocking curve from 0.4° to 0.5° due to the lattice constant relaxation (SI Appendix, Fig. S1C). According to polarization-electric field hysteresis loops the PMN-PT has a remnant polarization of 20 μC/cm$^2$ and a ferroelectric imprint of 50 kV/cm favoring the polarization pointing towards the SrRuO$_3$ (SI Appendix, Fig. S2A). The longitudinal piezoelectric response of the membrane was measured to be 1200 pm/V using a double-beam interferometer (SI Appendix, Fig. S3), comparable to that of 0.7PMN-0.3PT bulk single crystal samples (9).

A key aspect of our membranes is that the PMN-PT layer is continuous, with electrically biased regions (defined by patterned SrRuO$_3$ top electrodes) embedded in unbiased PMN-PT. A bias voltage applied between the continuous Pt bottom electrode and the patterned SrRuO$_3$ top electrode polarizes only this defined region of the PMN-PT, and we find that the intrinsic isotropic in-plane strain state is transformed by interaction with the surrounding unbiased PMN-PT into the anisotropic strain required to drive in-plane magnetic anisotropy. Anisotropic strain is present both inside and outside of the biased region, and the strain direction is spatially varying (Fig. 2B). We refer to this interaction as boundary clamping and show that it can be used to design an electric field induced strain that controls the in-plane magnetization orientation in the Ni regions. Our measurements of membrane composite magnetoelectrics show electric-field induced uniaxial anisotropy and are in good agreement with our analytical and numerical analyses of the piezoelectric strain tensor in this constrained geometry.

**Experimental Results**

Longitudinal magneto-optic Kerr effect (MOKE) magnetic hysteresis loops were used to measure strain-induced magnetic anisotropy in the Ni at different PMN-PT bias voltages. Applying the magnetic field along an easy magnetic direction will result in a square hysteresis loop as the magnetization jumps between orientations parallel and antiparallel to the applied field. Applied field along a hard direction rotates the magnetization away from the easy axis, resulting in a linear MOKE magnetic hysteresis loop with zero coercivity that saturates at an applied field $H_{\text{sat}}$. The uniaxial magnetic anisotropy energy density $K_U$ can be estimated from the hard axis data with $K_U = \frac{\mu_0}{2} M_s H_{\text{sat}}$, where $M_s$ is the Ni saturation magnetization, and assuming coherent rotation (20).

Figure 3A shows the bias dependence of MOKE hysteresis loops of a 300 μm by 200 μm Ni/SrRuO$_3$ rectangle that serves as top electrode for PMN-PT bias. In the top panel the applied field magnetic field is along $\hat{x}_1$, parallel to the long edge of a rectangle. As the applied bias is increased from 0V to 8V, the loops close from square to nearly linear, indicating the formation of a magnetic hard direction along $\hat{x}_1$ with an anisotropy energy of 1.2 kJ/m$^3$. In the bottom panel of Fig. 2A the measurement field is rotated by 90° to be along $\hat{x}_2$, parallel to the shorter edge of the pattern. As the applied bias increases, there is a small change in coercive field but no noticeable change in loop shape, showing that the $\hat{x}_2$ axis remains easy, independent of bias. These two measurements confirm that the piezoelectric strain has induced a new uniaxial anisotropy in the Ni layer along $\hat{x}_2$. As Ni has a negative magnetostriction constant, $\hat{x}_2$ must be the most compressively strained direction in the biased region. At zero bias, the hysteresis loops for both field directions (and all others measured but not shown) are identical, indicating no intrinsic anisotropy in this sample. Strain-induced uniaxial magnetic anisotropy is expected to immediately induce a hard-axis response, with a magnetic anisotropy proportional to strain, rather than the gradual loop closing observed experimentally. We believe that our experimental results arise from strain-induced anisotropy competing against domain wall pinning, consistent with the relatively large 10 mT Ni coercive field, attributable to growth conditions. Other devices showed the expected linear increase in anisotropy energy with increasing bias voltage (SI Appendix, Fig. S7).

Figures 3B and 3C show spatial maps of the Ni coercivity measured with MOKE. A complete hysteresis loop was measured with the laser focused at each 10μm x 10μm pixel and the magnetic field was aligned 30˚ from the previously determined strain-induced hard axis direction[1]. At zero bias (Fig. 3B), the coercive field is uniform and matches the zero-bias coercive field measured in Fig. 2A. At a 6V bias (Fig. 3C), the coercive field drops considerably indicating loop closure and a strain-induced magnetic ansiotropy. The loops don't close completely due to the 30° misalignment with the hard axis. The coercivity is lower near the center of the pattern, and higher near the short edges, suggesting a larger anisotropy near the center of the electrode, as expected based on our analysis below.

In addition to the strains within the biased region of the PMN-PT, there is also significant strain outside of the biased region. The strain-induced magnetic anisotropies inside and outside of the biased region are qualitatively different. To probe this difference, we patterned a device with a grid of 60 μm by 80 μm Ni islands, each free to respond independently to local strains, placed on and around a 300 μm by 200 μm $SrRuO_3$ electrode. MOKE magnetic hysteresis loops are shown for two nearby Ni islands at 0V and 5V, one inside (Fig. 4A) and one outside (Fig. 4B) of the biased region. Both islands have square hysteresis loops at zero bias[2] with the applied field along $\hat{x}_2$. At 5V, the Ni island inside the biased region has an unchanged hysteresis loop, matching the behavior of the larger Ni rectangle shown in Fig. 1A (bottom panel). The Ni island outside the bias region, under a 5V bias, develops an 0.84 kJ/m³ anisotropy parallel to the long edge and perpendicular to the anisotropy induced inside the biased region. This 90° difference in anisotropy is consistent with our detailed analysis presented below, but can also be understood qualitatively: when the biased PMN-PT compresses inwards, it stretches the unbiased region. The magnetization in the compressed region aligns parallel to the axis of compression (Fig. 4A, easy axis along $x_1$), and in the stretched region aligns perpendicular to the axis of tension (Fig. 4B, easy axis along $x_2$).

---

[1] Shadowing of the laser by a wire bond prevented measurement with the magnetic field along $x_1$ or $x_2$ so an intermediate angle was chosen.
[2] Due to different growth conditions, the Ni in the two islands in Fig. 4a and b have as-grown magnetic anisotropy along $\hat{x}_2$, in contrast to the Ni in the sample measured in Fig. 3 which was magnetically isotropic.

**Discussion**

Our magnetoelectric measurements demonstrate that piezoelectric strain is responsible for inducing, via magnetoelasticity (21), a magnetic easy axis along the shorter direction of rectangular electrodes. This would not occur without the boundary clamping of the biased PMN-PT by the surrounding unbiased PMN-PT. Here we develop an analysis that relates the piezoelectric strain, boundary clamping, and magnetic anisotropy, and which allows for the design of an electric field-dependent magnetic anisotropy pattern in the Ni layer through piezoelectric tensor design.

A bias applied across the thickness of the PMN-PT generates strain in the PMN-PT through the converse piezoelectric effect. Normal (non-shear) strains in cubic piezoelectrics are characterized by two piezoelectric tensor components, $d_{33} > 0$ and $d_{31} < 0$, which in this geometry respectively describe the elongation parallel and perpendicular to the applied electric field. Because of its tetragonal symmetry when polarized along $\hat{x}_3$, an unconstrained sheet of PMN-PT responds equally along $\hat{x}_1$ and $\hat{x}_2$ (i.e. $d_{31} = d_{32}$), creating isotropic strain. However, when only a small region of the membrane is biased, its contraction is constrained by the surrounding unbiased PMN-PT, resulting in anisotropic response. We find that the effect of this boundary clamping can be described with effective (subscript *eff*) piezoelectric tensor components of the biased region, with $d_{32,eff} \neq d_{31,eff}$. This modification leads to strain-dependent uniaxial magnetic anisotropy.

This magnetic anisotropy induced by the applied bias depends on the in-plane components of the strain tensor, which are spatially varying due to the boundary clamping. Locally, every two-dimensional strain distribution has a direction of maximum strain and minimum strain, referred to as the first and second principal strain directions.[3] The notation $\varepsilon_1$ and $\varepsilon_2$ is used here to denote the magnitude of the first and second principal strains. In terms of the principal strains, the induced anisotropy energy in the presence of an arbitrary strain distribution is

$$K_U = -\frac{3}{2} \lambda_S Y_{Ni} (\varepsilon_1 - \varepsilon_2) \qquad (1)$$

where $\lambda_S$ and $Y_{Ni}$ are the saturation magnetostriction constant (-32.9 ppm) and Young's modulus (220 GPa) of polycrystalline Ni (21). This means that however complex the strain distribution, it locally induces a uniaxial anisotropy, with direction and magnitude determined by the principal strains of the strain tensor. Here the anisotropy axis is parallel to the second principal strain direction, because this is the most compressed direction and Ni has a negative $\lambda_S$. Figure 2B schematically shows the principal strains at three infinitesimal regions in the biased and unbiased regions of a piezoelectric membrane. Upon applying a bias, the gray (undeformed) square patches are stretched or compressed into the black rectangular patches, each with its own principal strain directions.

We can estimate the strain difference $\varepsilon_1 - \varepsilon_2$ at the center of the biased region from (1) using the values of $K_U$ from our MOKE hysteresis loops. Considering only strain-induced anisotropies, the hard axis measurement in Fig. 3A gives $\varepsilon_1 - \varepsilon_2 = 110$ ppm at 8 V bias, and that of Fig. 4B gives $\varepsilon_1 - \varepsilon_2 = 78$ ppm at 5V bias. The effective piezoelectric constants may also be estimated as the strain difference per applied electric field, giving $d_{31,eff} - d_{32,eff} = 6.9$ pm/V.

---

[3] The magnitude and direction of the principal strains are the eigenvalues and eigenvectors of the strain tensor. In the case of isotropic strain, the eigenvalues are degenerate, and no direction of maximum strain exists. Shear strain terms vanish upon coordinate transformation into the frame defined by the principal strain directions.

**Analysis**

We find that strain patterns in our piezoelectric membranes can be understood by building on a continuum elasticity theory first developed by Eshelby (22) to describe the elastic behavior of precipitates in materials. An exactly ellipsoidal region embedded in an elastic media will strain anisotropically in response to an isotropic internal stress, with the strain exactly uniform inside the ellipsoid. The strain is largest along the shortest axis of the ellipsoid. This is in agreement with our experimental results: the biased regions in our samples undergo uniform stress from their piezoelectric response, and our MOKE measurements indicate that the largest compressive strain lies along the shorter axis of rectangular patterns, in agreement with Eshelby's model.

Inside an infinite elliptical cylinder with axes $a$ and $b$, respectively along $\hat{x}_1$ and $\hat{x}_2$, the strain response to an electric field along the cylinder axis is (23)

$$\varepsilon_{ij} = E_3 \frac{e_{31}}{(a+b)\, c_{11}} \begin{pmatrix} b & 0 \\ 0 & a \end{pmatrix} = E_3 \frac{e_{31}}{(1+A)\, c_{11}} \begin{pmatrix} 1 & 0 \\ 0 & A \end{pmatrix} \qquad (2)$$

where $\varepsilon_{ij}$ is the strain tensor, $E_3$ is the electric field, $e_{31}$ is the transverse piezoelectric coupling constant ($e_{ij} = c_{ik}\, d_{kl}$), and the aspect ratio $A = \frac{b}{a}$. The resulting first and second principal strains are $\varepsilon_{11}$ (along $\hat{x}_1$) and $\varepsilon_{22}$ (along $\hat{x}_2$). The magnetic anisotropy induced by this strain distribution, as a function of aspect ratio and applied electric field, is found from equations (1) and (2) to be

$$K_U = -\frac{3}{2} \lambda_S\, Y_{Ni}\, E_3 \frac{e_{31}}{c_{11}} \frac{1-A}{1+A}. \qquad (3)$$

Using bulk materials constants (24) in this model yields $K_U = 1.1$ kJ/m$^3$ for an 8 V bias across a 3:2 aspect ratio ellipse, close to the measured value 1.2 kJ/m$^3$ for our rectangular electrodes. This order of magnitude agreement suggests that far inside the pattern edges the aspect ratio primarily determines the effect of boundary clamping on the electric field induced magnetic anisotropy. The magnitude of the magnetic anisotropy is independent of the absolute size of the biased region, suggesting that lateral electrode dimensions much smaller than the 100 μm scale used here would still be effective.

Finite element continuum elastic simulations were performed to address the rectangular biased regions used in our experiments, mapping strains and the resulting magnetic anisotropy (Fig. 4C). All layers of the structure shown in Fig. 2A except the Au and SU-8 were included in the simulation, using bulk values for the elastic, piezoelectric and dielectric tensors of PMN-PT (24). Figure 4C shows the strain-induced magnetic anisotropy energy per applied voltage (color) and anisotropy axis (white lines) on the surface of the PMN-PT layer. The computed anisotropy predominantly perpendicular and parallel to the long edge inside and outside the biased region respectively reproduces the experimental results of Fig. 4A and 4B. The change in direction near the short edge coincides with very small anisotropy magnitude, and so is difficult to detect experimentally. The computed 0.45 kJ m$^{-3}$V$^{-1}$ magnitude in the large central portion of the biased region predicts a 3.6 kJ/m$^3$ anisotropy energy density at 8V, about three times the experimental value. The computed anisotropy is largest near the center, consistent with the experimental spatial maps of Fig. 3C.

We also simulated a series of elliptical electrodes with varying aspect ratios to compare with the Eshelby approach. Figure 4D shows that the simulated and analytical anisotropy energies have the same $\frac{A-1}{A+1}$ dependence on aspect ratio. However, the analytic result of Eq. (3) describe an infinite cylinder of PMN-PT, and the simulation the experimental two-dimensional composite sheet. The two *y*-axis scales in Fig. 4D indicate that this difference results in different predicted anisotropy energy magnitudes, but the dependence on aspect ratio is captured by the analytic result.

According to finite element calculations, the area of largest uniaxial strain is just outside of the biased region boundary (Fig. 4C). Analytical solutions (25) for the strain outside of elliptical precipitate inclusions confirm that the largest uniaxial strains are concentrated on the most curved portion of the boundary, and that the strains drop off like $1/|x|^2$ far from the boundary (SI Appendix, Note 1). The measured anisotropy directions of the Ni island outside the biased region (Fig. 4B) match the calculated anisotropy parallel to the long edge (Fig. 4C black boxed) and the direct phase-field simulations of the magnetization direction (SI Appendix, Fig. S6) for those locations. We did not experimentally find a significant difference in the induced anisotropy energy inside and outside of the biased region, likely due to pre-existing magnetic anisotropy and domain pinning in the Ni that makes $H_{\text{sat}}$ a coarse method for measuring anisotropy energy.

**Conclusions**

The preceding analysis, and that summarized in the SI Appendix, leads to a set of guidelines for manipulating the magnetoelectric response in piezoelectric membrane composites using piezoelectric tensor design. An elongated single electrode generates uniaxial compressive strain and magnetic anisotropy inside the biased piezoelectric region that increases with aspect ratio, and is predominantly oriented along the short axis. Ellipsoidal biased regions have exactly uniform interior strains (SI Appendix, Fig. S4), with about sixty percent of the limiting anisotropy value obtained at an aspect ratio of 4:1. Substantial further increases require large increases in aspect ratio. Rectangular regions generate about 20% more uniaxial strain than ellipses of the same aspect ratio (SI Appendix, Fig. S5), but the strain is less uniform in rectangles. The maximum uniaxial tensile strain is located outside highly curved boundaries and is at least twice as large as the interior uniaxial strain, but at the cost of reduced spatial uniformity (SI Appendix, Note 1). In the case of a straight boundary, the exterior magnetic anisotropy is perpendicular to the interior anisotropy. These rules allow for the design of particular anisotropy magnitudes and directions using boundary shape and layout.

Piezoelectric membrane composites are positioned to take advantage of interest in freestanding films and the number of available fabrication techniques (26, 27). Our results here demonstrate the fundamental principles of piezoelectric tensor design for magnetoelectric coupling in membrane composites, and optimization of the biased region geometry will likely realize even higher magnetoelectric coupling. Several theoretical proposals for inducing 180° in-plane magnetization rotation in bulk composite magnetoelectrics using spatially varying electric fields have been proposed (28–30). Piezoelectric membranes offer an alternative, thin-film platform for realizing such proposals using the design guidelines developed here. We also expect that these magnetoelectric membrane structures can be used as sensors, with a piezovoltage readout. Although we have focused on in-plane magnetization manipulation, the biaxial strain present in square or circular devices may also be able to control the out-of-plane magnetization of a ferromagnetic overlayer with perpendicular magnetic anisotropy (31, 32). Additionally, integration of other materials with piezoelectric membranes would allow for electric field control of, for instance, superconducting $T_C$ (33–35), band topology (36–38), conductivity (39), and optical properties (40) with designed strain patterns.

## Materials and Methods

Membrane Fabrication

Figure 1 is the schematic of the fabrication procedure for the PMN-PT membrane devices. Here we will describe the method in detail. Growth of high quality epitaxial PMN-PT (001) thin films started with a (001) Si wafer with a 4° miscut towards [110] direction and a 20nm buffer layer of STO. First, 100 nm of SRO was grown using 90° off-axis rf-magnetron sputtering (41) at 100 W power and 600 ºC. A mixture of Ar and $O_2$ gas, flown at 12 sccm and 8 sccm respectively, supplied a working pressure of 200mTorr. PMN-PT films were then grown using a misaligned parallel dual planar magnetron sputtering technique (42) with substrate rotation with 100W power at 625 ºC. A mixture of Ar and $O_2$ gas, flown at 17 sccm and 3 sccm respectively, supplied a working pressure of 500 mTorr for PMN-PT growth. A 100 nm layer of Pt was then deposited on top of PMN-PT by DC Magnetron sputtering. The heterostructure was annealed in $O_2$ at 300 ºC for 30 minutes to reduce residual stress in the Pt film. The Si substrate was then mechanically polished to reduce the thickness from 300 μm down to 100 μm to reduce total etching time during the later $XeF_2$ dry etching. After polishing, Polydimethylsiloxane (PDMS), with a 10:1 mixture ratio of monomer to crosslinking agent, was spin-coated onto a glass slide at 5000rpm for 10 seconds, resulting in a thickness of approximately 30μm. The thin film heterostructure was then placed Pt-side down onto the uncured PDMS, leaving the Si substrate exposed, and the sample was placed under vacuum for a minimum of 5 hours to remove air bubbles from between the Pt and PDMS layers. After the vacuum treatment, the PDMS was then cured on a hot plate for 1 hour at 100 ºC.

Once the PDMS was cured, the sample was ready for Si removal. Prior to $XeF_2$ etching of Si, a 15 second plasma etch using $CF_4$ and $O_2$, flowing at 45 sccm and 5 sccm, respectively, to a pressure of 40mTorr, was performed to remove any moisture on the sample, as well as any native SiO$_2$ present on the Si substrate. The Si substrate was then completely removed via XeF$_2$ etching. The XeF$_2$ etching system was performed in a SPTS Xetch e1 XeF$_2$ etcher system. The system exposes the samples to XeF$_2$ in a cyclic mode, and the recipe used here was chosen to maximize the etch rate for complete removal of the Si substrate. The recipe exposed the samples to 4 Torr of XeF$_2$ for two-minute periods, followed by pumping down to 0.8 Torr between cycles for a continuous etch before the next cycle began. Due to the exothermic nature of the reaction of XeF$_2$ with Si, the pressure in the chamber rises during the two-minute etch cycles. When the Si is completely removed, the pressure increase is notably absent during an etch cycle, signaling that the etching is complete.

After Si removal, the STO buffer layer was removed via Ar$^+$ ion-milling. The SRO layer was then patterned into various geometries using photolithography and wet etching with a 0.4M NaIO$_4$ solution. A 35 nm layer of Ni was deposited by DC Magnetron sputtering, and photolithography was performed to pattern the Ni with a Transene Ni Etchant Type 1 wet etchant. A SU-8 protection layer was applied by spin-coating at 5000 rpm for 40 s, resulting in a thickness of 2μm, followed by photolithography patterning. Finally, 30 nm of Au was deposited via DC Magnetron sputtering and patterned via photolithography, and Transene TFA Au wet etchant was used to make the "lifted" Au electrodes.

Finite-Element Simulations

Finite element calculations were performed with COMSOL Multiphysics$^{TM}$. Simulations were performed using the layers and thicknesses from Figure 2A. The sheet of PMN-PT and back electrode were 1.4 μm diameter to simulate a small biased region surrounded by a large unpolarized membrane. No mechanical constraints were applied to any surfaces, simulating an unconstrained membrane. The stiffness tensor and piezoelectric coefficients used for PMN-PT may be found in Table 2 of reference (24). The stress-charge form of the piezoelectric constitutive relations was used:

$$\nabla \cdot D_i = \rho_f$$

$$\nabla \cdot \sigma_{ij} = 0$$

$$D_i = e_{ikl}\varepsilon_{kl} + \varepsilon_0 \kappa_{ij} E_j$$

$$\sigma_{ij} = c_{ijkl}\varepsilon_{kl} - e_{kij}E_k$$

where $D_i$, $E_i$, $\sigma_{ij}$, $\varepsilon_{kl}$, $\rho_f$, $e_{ijk}$, $c_{ijkl}$, $\kappa_{ij}$ are the electric displacement, electric field, stress tensor, strain tensor, free charge density, piezoelectric coupling tensor, stiffness tensor and relative permittivity, respectively.

Longitudinal MOKE Measurements

The sample was mounted between the poles of an electromagnet and a red HeNe (632nm) laser was reflected off of the sample surface at approximately 45° from normal incidence. The beam was focused to an approximately 10 μm spot using an achromat. The reflected beam's polarization was rotated to 45° from *p*-polarized with a half-wave plate and then the *s*- and *p*-polarized components were measured with a differential balanced photodetector. The differential signal is proportional to the Kerr polarization rotation. Spatial mapping was achieved by mounting the sample on a two-axis linear piezoelectric motion stage and scanning the sample under the focused beam.

**Acknowledgements**
This work was supported by the Army Research Office through grant W911NF-17-1-0462.

**Author Contributions**
J.I., S.L., W.M., M.S.R. and C.B.E. designed research; J.I., S.L., V.V. and J.W. performed research; J.I., S.L., L.C., J.H, M.S.R. and C.B.E. analyzed data; J.I., S.L., M.S.R. and C.B.E. produced the manuscript.

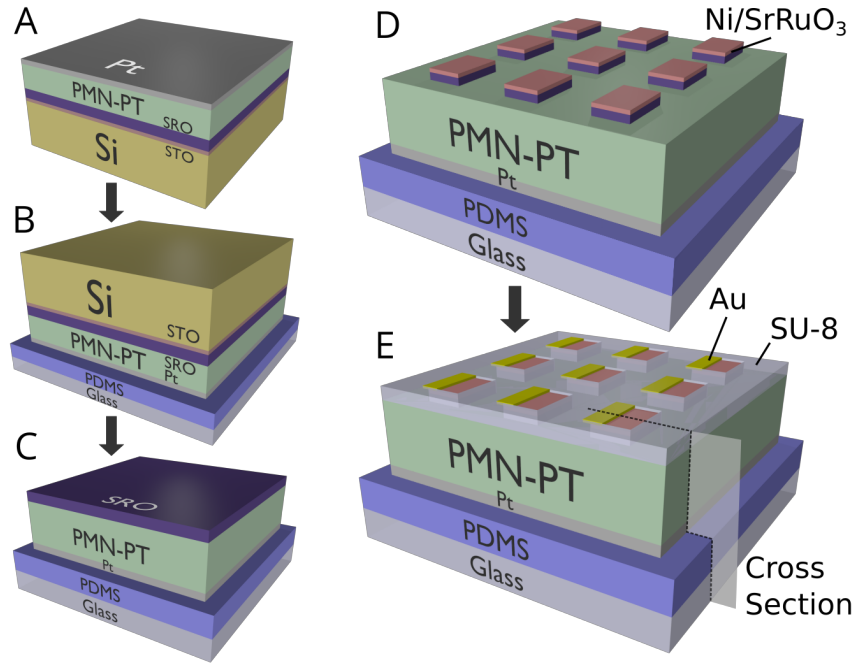

**Fig. 1** Schematic of the fabrication procedure for membrane magnetoelectric devices. (A) PMN-PT/SrRuO$_3$/SrTiO$_3$/Si thin-film heterostructure with Pt electrode. (B) Heterostructure is flipped an attached to PDMS coated glass. (C) Si and SrTiO$_3$ (STO) are etched off leaving behind sub-micron membrane. (D) Ni is deposited and Ni/SrRuO$_3$ (SRO) is patterned into an array of devices. (E) A protective coating of SU-8 is applied and Au contacts are added. The indicated cross section view is shown in Fig. 2A.

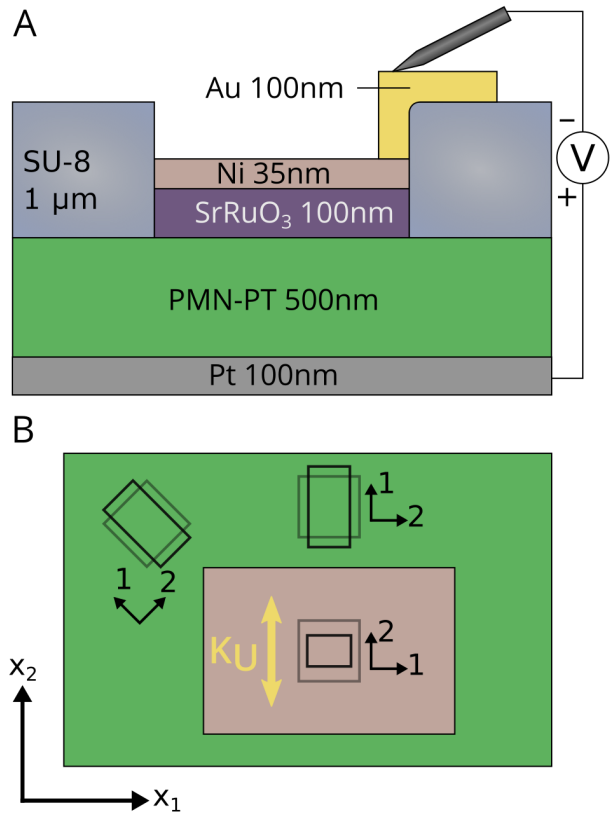

**Fig. 2.** (A) Cross section schematic of a completed sample. Positive voltage corresponds to polarization towards the SrRuO$_3$. (B) Biased regions with aspect ratio A $\neq$1 generate excess strain along their shorter directions which induces a magnetic easy direction (gold arrow). The strain distribution in both the biased (gray) and unbiased (green) regions can be considered in terms of local principal strains, shown with small axes indicating directions of principal strains $\varepsilon_1$ and $\varepsilon_2$. The grey squares represent undeformed infinitesimal patches and the black rectangles represent the same patches after deformation due to the piezoelectric response.

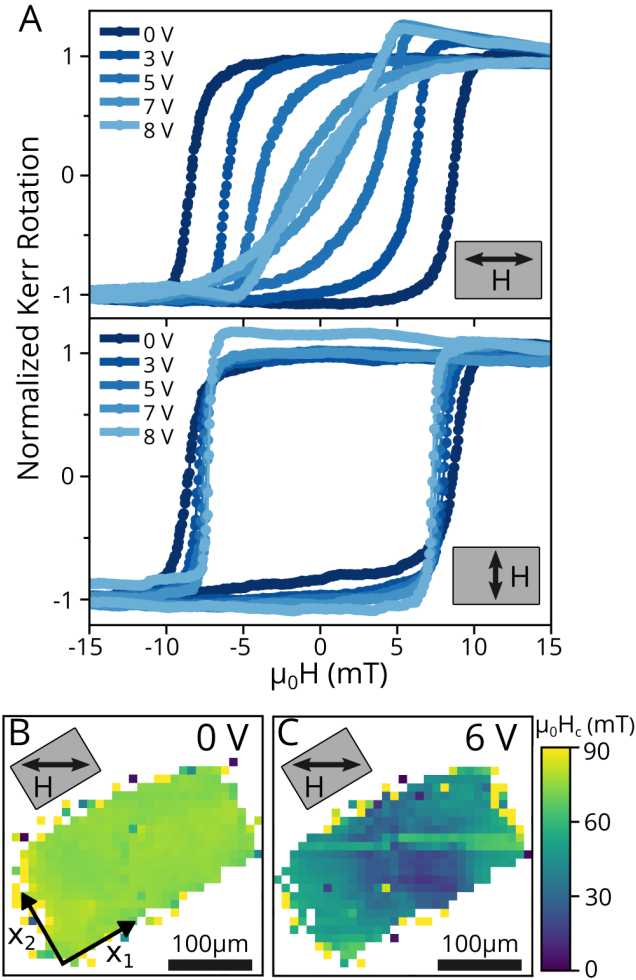

**Fig. 3.** (A) Magnetic hysteresis loops with the applied magnetic field parallel (top panel) and perpendicular (bottom panel) to the long edge of the pattern. (B) Map of coercive field ($H_C$) across a Ni island measured with MOKE at zero bias. No magnetic signal was detected at white pixels. (C) Coercive field map of the same island with a 6V applied bias.

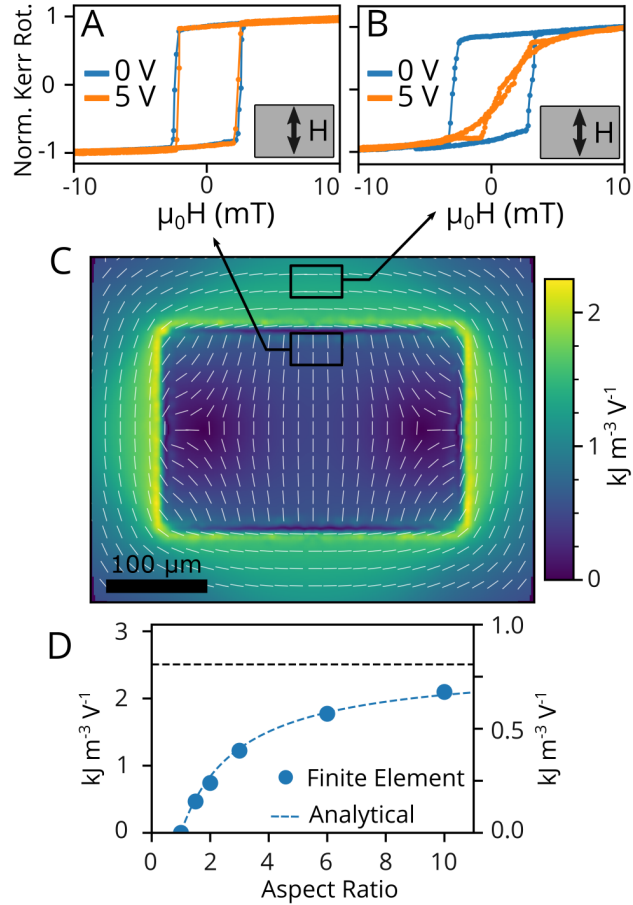

**Fig. 4.** MOKE hysteresis loops measured on Ni islands placed just inside (A) and outside (B) a 3:2 biased region at 0 V and 5 V bias and with the magnetic field along $\hat{x}_2$. (C) Induced magnetic anisotropy per applied voltage on and around the biased region. Color represents anisotropy energy and the white lines are the anisotropy axis direction. Black rectangles indicate the experimentally probed regions. (D) Aspect ratio dependence of the simulated (circles, left axis) and Eshelby model (dotted lines, right axis) anisotropy energy inside elliptical biased regions. The asymptotic value is shown as a dashed black line.

# Supplementary Information for

Magnetoelectric Coupling by Giant Piezoelectric Tensor Design

J. Irwin, S. Lindemann, W. Maeng, J. J. Wang, J.M. Hu, L.Q. Chen, D.G. Schlom, C.B. Eom, M.S. Rzchowski

M.S. Rzchowski
Email: rzchowski@physics.wisc.edu

**This PDF file includes:**

    Material Characterization
    Modeling of Piezoelectric Membrane Devices
    Supplemental Magnetoelectric Characterization
    Figs. S1 to S7
    Table S1
    Note 1

# Material Characterization

Structural Characterization of Membrane Devices

The structure of the PMN-PT thin films was measured by a high-resolution four-circle XRD machine (Bruker D8 Discover). **Figure S1A** shows a θ-2θ scan for a PMN-PT thin film before and after substrate removal. Before substrate removal the heterostructure measured here was as follows: 500 nm PMN-PT /100 nm SrRuO$_3$ / 20 nm SrTiO$_3$/ Si (001). Therefore, observed peaks include peaks from the Si, SRO, and PMN-PT layers, but the STO layer was too thin for peaks to be observed in this measurement. Once the membrane device was completed, the heterostructure measured was: Ni / SRO / PMN-PT / Pt / PDMS / glass. The SU-8 and Au were not yet deposited onto this device when it was measured. The SRO and Ni layers are patterned into small features of 40-200 μm with large spaces in between, therefore, the SRO peaks are greatly reduced in intensity. The major observed peaks only stem from the PMN-PT and Pt layers.

As expected, the Si 004 peak disappears after Si has been etched away, and the SRO peaks also vanish after SRO has been patterned, but one other noticeable change is that the PMN-PT peaks shift towards lower 2θ values meaning that the out-of-plane lattice parameter has increased due to the lattice mismatch strain relaxation. **Figure S1B** shows only the region around the (002) peaks, and a reference line has been added to highlight the difference in the PMN-PT peak position once the Si is removed. **Table S1** shows the 2θ values and corresponding lattice spacings for the PMN-PT thin film, membrane, and bulk PMN-PT with a composition near the MPB for comparison. Once the membrane has been freed from the Si substrate, the 2θ decreases towards the bulk PMN-PT value, showing that the PMN-PT film grows under tensile in-plane strain due to lattice parameter mismatch with Si substrate and relaxes back towards its bulk value upon substrate removal. An increase in the width of the 2θ peaks accompanies this relaxation.

**Figure S1C** shows the rocking curves of the PMN-PT 002 peaks before and after substrate removal. For the PMN-PT on the Si substrate, the FWHM = 0.4º. Once the PMN-PT is released and relaxes the FWHM = 0.5º, a slight increase. **Figure S1D** shows phi scans of PMN-PT 101 peaks for the Film and Membrane devices. The FWHM of the phi scans for the film and membrane are 0.95º and 0.99º, respectively.

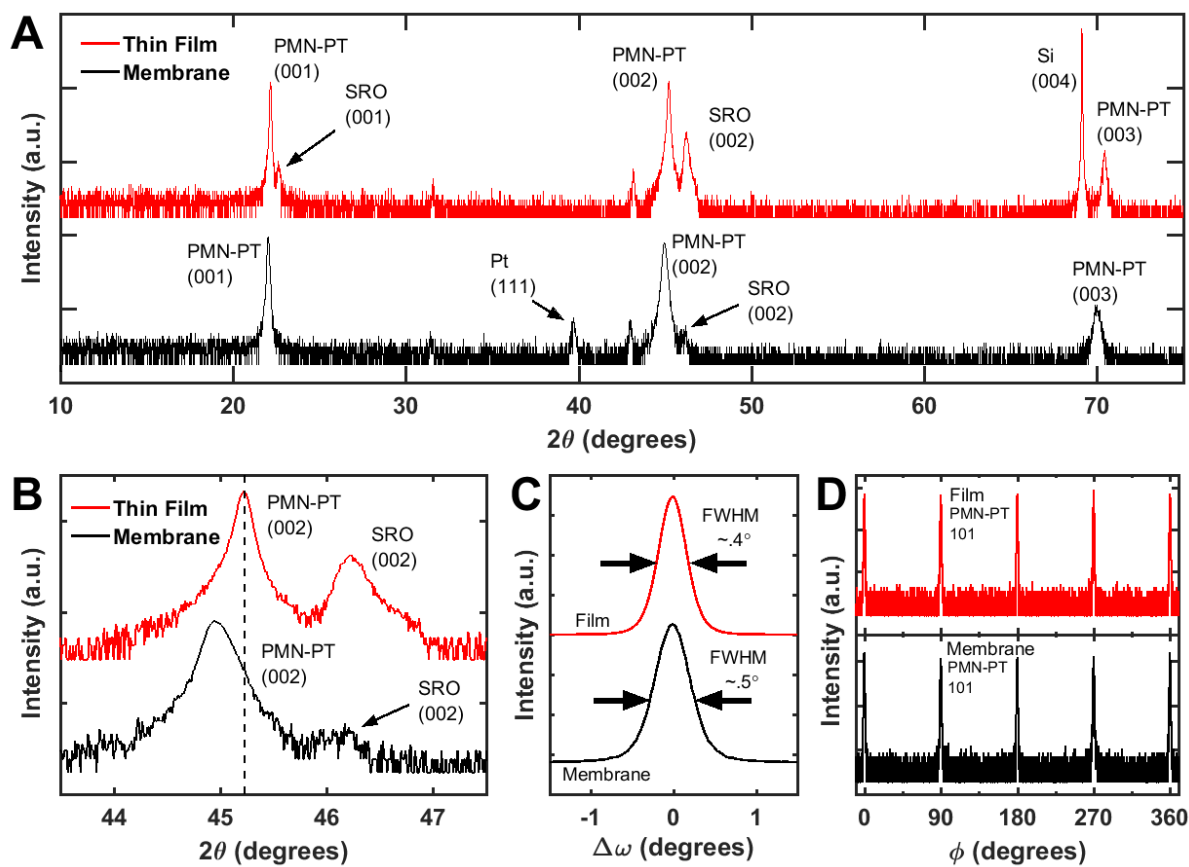

**Fig. S1.** X-ray Diffraction Characterization of PMN-PT Thin Film and Membrane (A) Full θ-2θ XRD scan of PMN-PT Thin Film and Membrane. (B) θ-2θ scan region around the (002) PMN-PT peak shows a shift in 2θ for the PMN-PT peak once the Si substrate is removed. The shift demonstrates the tendency for the PMN-PT to relax towards bulk PMN-PT lattice spacings (**Table S1**). The membrane device was measured after SRO patterning, therefore the SRO peaks have very little intensity. (C) Rocking Curve of (002) PMN-PT Thin Film Peak and Membrane. There is a slight increase in FWHM upon substrate release. (D) Phi scans of (101) PMN-PT Thin Film and Membrane.

## Table S1. Out-of-plane Lattice Spacing of PMN-PT

| Sample | 2θ | d-spacing |
|---|---|---|
| Thin Film | 45.225° | 4.0068 Å |
| Membrane | 44.960° | 4.0292 Å |
| Bulk* | 44.800° | 4.0428 Å |

*Bulk Value for PMN-PT Compositions near the MPB: (x)PMN-(1-x)PT x ≈ 0.3

Ferroelectric and Dielectric Characterization of PMN-PT Membrane

**Figure S2A** compares polarization-electric field hysteresis loops of a PMN-PT thin film to a membrane released from its substrate. Both film and membrane show ferroelectric imprint that favors polarization pointing towards the top ($SrRuO_3$) electrode. The membrane appears to have a slightly larger saturation polarization, but this could also be an artifact due to increased leakage currents during the measurement. **Figure S2B** shows that the dielectric constant of the membrane in the polarization reversal region is smaller in the thin-film sample, probably due to substrate clamping.

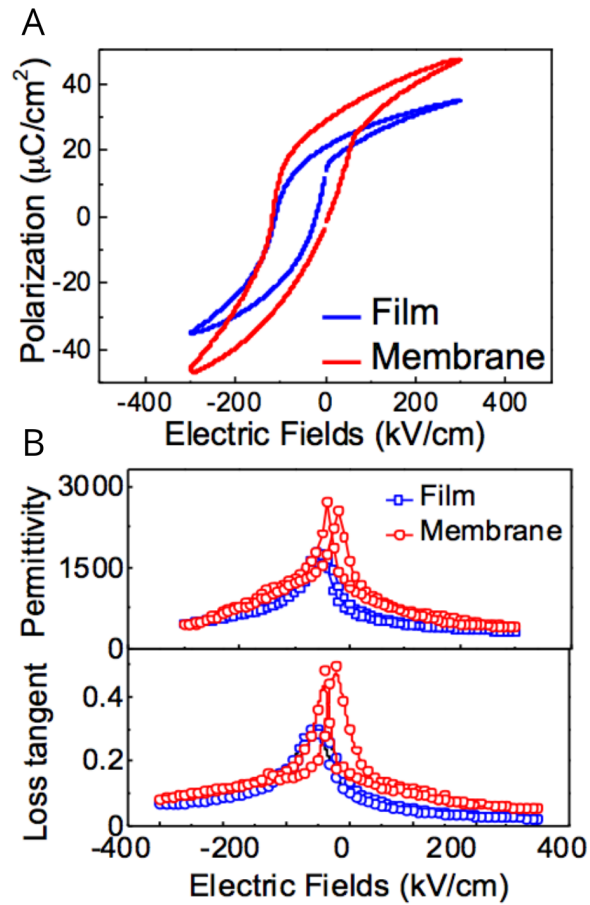

**Fg. S2.** Ferroelectric and Dielectric Characterization of Films and Membranes
(A) Polarization-Electric field loops for thin-film (blue) and membrane (red) samples. (B) Permittivity and Loss Tangent measurements of thin-film (blue) and membrane (red) samples.

Piezoelectric Characterization of PMN-PT Membrane

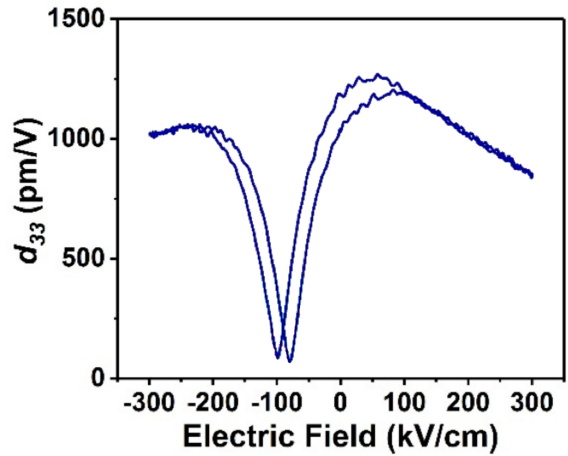

**Fig. S3.** Piezoelectric $d_{33}$ measured with a double beam intereferometer.

# Modeling of Piezoelectric Membrane Devices

Anisotropy Map of 2:1 Elliptical Biased Region

**Figure S4** shows a finite-element simulation of an elliptical biased region using the same methods and materials parameters as Figure 3 shows that the interior uniaxial strain is exactly uniform, as predicted by the Eshelby model. Outside the biased region the anisotropy direction is approximately tangent to the nearest boundary point. The anisotropy direction does not necessarily rotate by 90° upon crossing from the biased to unbiased region, for example near the biased region boundary on the major axis. The rectangular biased region simulation of Figure 3 demonstrated the 90° rotation when crossing both the long and short boundaries, although the anisotropy strength was very low near the short edge, and the direction did not rotate exactly 90° near the corners of the rectangle.

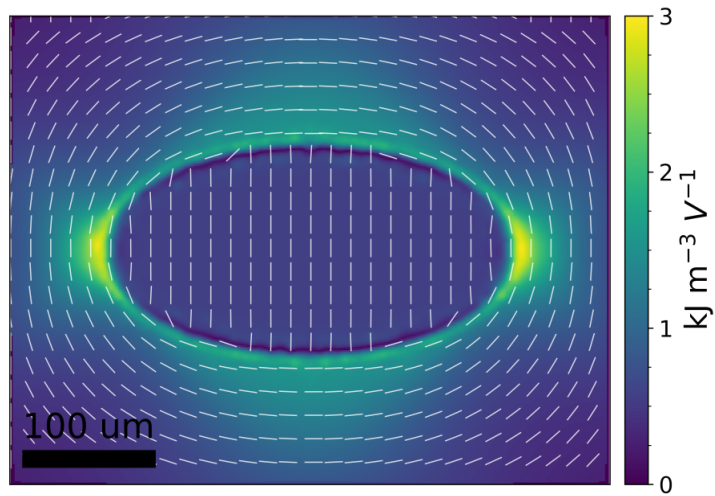

**Fig. S4.** Finite-element simulation of a 300 μm by 150 μm elliptical biased region.

## Strain in Elliptical vs. Rectangular Biased Regions

Finite-element simulation found that the strain in the central region of the biased region is consistently larger in rectangular biased regions than in elliptical ones (**Fig. S5**). The net effect of the curvature of the elliptical boundaries is to generate uniform interior strain, but at the cost of about a 20% reduction in strain magnitude. Rectangular electrodes have non-uniform strain away from their centers, but the strain per aspect ratio, and per electrode area, is greater.

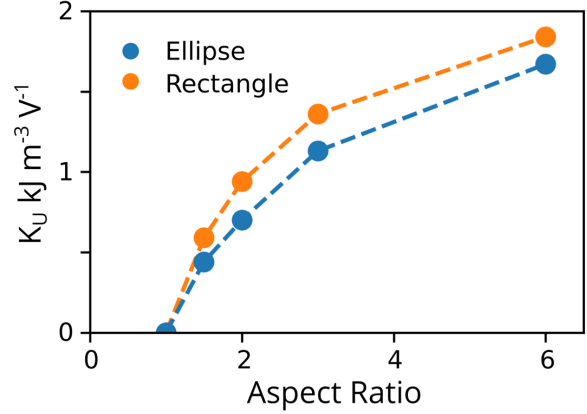

**Fig. S5.** Comparison of the strain-induced magnetic anisotropy strength per volt at the middle of elliptical and rectangular biased regions, computed with finite-element simulations.

## Phase-field Method

In the phase-field model of magnetics, the magnetization is selected as the order parameter to describe the magnetic anisotropy and magnetic domain structures, which can be evolved by the Landau–Lifshitz–Gilbert equation

$$(1+\alpha^2)\frac{\partial \mathbf{M}}{\partial t} = -\gamma_0 \mathbf{M}\times\mathbf{H}_{eff} - \frac{\gamma_0 \alpha}{M_S}\mathbf{M}\times(\mathbf{M}\times\mathbf{H}_{eff}), \tag{S6}$$

where $M_S$, $\alpha$, and $\gamma_0$ represent the saturated magnetization, damping constant, and gyromagnetic ratio, respectively. The effective field is given by

$$\mathbf{H}_{eff} = -\frac{1}{\mu_0}\frac{\delta F}{\delta \mathbf{M}} \tag{S7}$$

with $F$ the total free energy including the magneocrystalline anisotropy energy, exchange energy, magnetostatic energy, external magnetic field energy, and elastic energy

$$F = F_{anis} + F_{exch} + F_{ms} + F_{external} + F_{elastic}. \tag{S8}$$

In the simulation, the $F_{anis}$ is neglected for simplicity regarding the isotropic nature of the polycrystalline Ni thin film. The isotropic $F_{exch}$ is determined by the gradient of local magnetization vectors, i.e.,

$$F_{exch} = \int_V \zeta\left[(\nabla m_1)^2 + (\nabla m_2)^2 + (\nabla m_3)^2\right]dV, \tag{S9}$$

where $\zeta$ denotes the exchange stiffness constant.

The magnetostatic energy density $f_{ms}$ can be written as,

$$F_{ms} = -\int_V \frac{1}{2}\mu_0 M_S (\mathbf{H}_d \cdot \mathbf{m}) dV. \tag{S10}$$

Here $\mathbf{H}_d$ denotes the stray field, and it can be numerically calculated by employing a finite-size magnetostatic boundary condition previously developed for a 3D array of ferromagnetic cubes (1).

The Zeeman energy of an external magnetic field can be expressed as

$$F_{external} = -\int_V \mu_0 M_S (\mathbf{H}_{ext} \cdot \mathbf{m}) dV. \tag{S11}$$

The elastic energy $F_{elastic}$ is written as

$$F_{elastic} = \frac{1}{2}\int c_{ijkl}(\varepsilon_{ij} - \varepsilon_{ij}^0)(\varepsilon_{kl} - \varepsilon_{kl}^0) dV \tag{S12}$$

through which the magnetoelastic coupling within the Ni island is considered. The total strain $\varepsilon_{ij}$ includes a homogeneous part $\overline{\varepsilon}_{ij}$ and an inhomogeneous part $\delta\varepsilon_{ij}$, which can be solved from the mechanical equilibrium equation. The homogeneous strain is assumed to be equal to the in-plane average piezoelectric strain at the surface region of the biased PMN-PT substrate underneath the Ni island.

To solve the phase-field equations of Ni nanoislands, spectral-based approaches are employed with following material parameters for Ni nanoislands: $M_s$=2.9×10$^5$ A/m (2), $\gamma_0$=2.2×10$^5$ m/(A·s) (3), $\alpha$=0.1 (3), $\lambda_s$=−3.3×10$^{-5}$ (2), $c_{11}$=247 GP (4), $c_{12}$=147 GP (4), $c_{44}$=50 GP (4), $\zeta$ =8.2×10$^{-12}$ J/m (5). The discrete grid points of 800$\Delta x$×600$\Delta y$×$\Delta z$ with real grid spaces $\Delta x$=$\Delta y$=5 nm. The real time step $\Delta t$ of 0.1 ps is used for solving the LLG equation.

**Figure S6A** shows the magnetization domain structures for Ni islands with zero bias across the PMN-PT membrane, showing multi-domain states of different orientations, indicating that there is no magnetic anisotropy. **Figure S6B** shows the anisotropic strain distribution computed by the phase-field method, with an 8V bias applied to the 300 µm by 200µm rectangular top electrode. The strain anisotropy transferred to 4 µm by 3 µm Ni islands on top of the membrane depends on their positions relative to the biased region. At position 1, as shown in **Fig. S6C**, the Ni island magnetization will be switched from multi-domain to single domain with the magnetization parallel to the short edge direction. Similarly, for a Ni island grown at position 2, the simulated magnetization is parallel to the long direction (**Fig.S6D**), indicating a magnetic easy axis in this direction. The phase-field direct simulation of the magnetoelastic coupling effect on the magnetization agrees with the experimental magnetic easy axis determination from Kerr magnetic hysteresis loops.

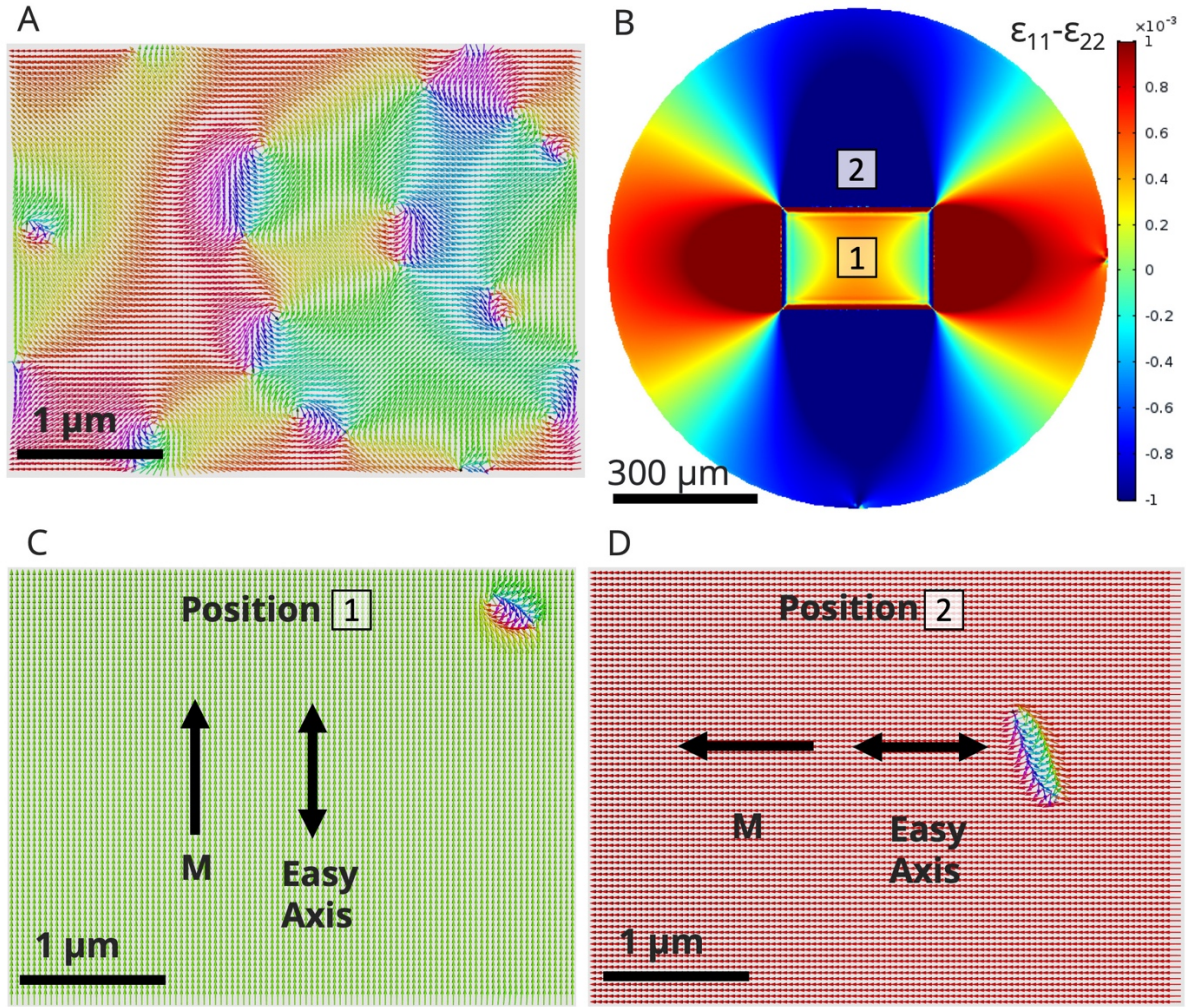

**Fig. S6.** Micromagnetic simulations of piezoelectric membrane device. (A) Magnetization domain structures for as-grown 4 µm by 3 µm Ni islands on PMN-PT without an applied electric field. (B) Anisotropic strain distribution used for micromagnetic simulations with Ni island positions labelled. Magnetization domain structures for Ni islands located at position 1 (C) and position 2 (D) with 8V bias voltage.

Analysis of Exterior Strain

**Note 1**

In this section we will provide analytic support for the design rules relating to magnetic anisotropy inside and outside elliptical biased regions. Jaswon and Bhargava (6) developed an analysis of the elastic response of two-dimensional elliptical elastic inclusions in an elastically isotropic media. Here we use this approach to investigate elliptical biased regions in a piezoelectric membrane, assuming isotropic elastic constants for simplicity. For an elliptical biased region with long axes $a$ along $x_1$ and short axis $b$ along $x_2$, confocal elliptical coordinates $(\xi, \eta)$ with foci a$\pm\sqrt{a^2 - b^2}$ are used to describe the exterior strains. The transformation to Cartesian coordinates is

$$\begin{cases} x_1 = \sqrt{a^2 - b^2} \cosh \xi \cos \eta \\ x_2 = \sqrt{a^2 - b^2} \sinh \xi \sin \eta \end{cases} \quad (S1)$$

Following Jaswon, the exterior strain can be written

$$\varepsilon_{\xi\xi} = -\varepsilon_{\eta\eta} = \gamma \frac{ab}{a^2 - b^2} \frac{e^{2\xi} - \cos 2\eta}{1 + e^{4\xi} - 2e^{2\xi} \cos 2\eta} \quad (S2)$$

$$\gamma = E_3 \, d_{31}(1 + \nu)$$

where $\nu$ is the shear modulus of the membrane (the expressions from the reference have been converted into strains using in the limit of plane stress). Along the $x_1$ axis $\eta = 0$ and along the $x_2$ axis $\eta = \pi/2$. For these two high symmetry directions the elliptical $\xi$ and $\eta$ directions align with the Cartesian axes making the induced anisotropy energy

$$K_U = -\frac{3}{2} \lambda_S \, Y_{Ni} |\varepsilon_{\xi\xi} - \varepsilon_{\eta\eta}| = -\frac{3}{2} \lambda_S \, Y_{Ni} |\varepsilon_{xx} - \varepsilon_{yy}| = -3\lambda_S \, Y_{Ni} \, \varepsilon_{\xi\xi}.$$

The boundary of the elliptical biased region is at $\xi = \xi_0$, with $\xi_0$ determined from any of the following equivalent relations:

$$\cosh 2\xi_0 = \frac{a^2 + b^2}{a^2 - b^2}, \quad \tanh \xi_0 = \frac{b}{a}, \quad e^{2\xi_0} = \frac{a+b}{a-b}. \quad (S3)$$

The maximum and minimum anisotropy energies occur on the $\xi = \xi_0$ contour at $\eta = 0$ and $\eta = \pi/2$, respectively. Combining (S2) and (S3) and converting to anisotropy energy gives

$$K_U|_{\eta=0,\xi=\xi_0} = K_U^{max} = -3\lambda_S \, Y_{Ni} \, \gamma \frac{A}{A+1} \quad (S4)$$

$$K_U|_{\eta=\pi/2,\xi=\xi_0} = K_U^{min} = -3\lambda_S \, Y_{Ni} \, \gamma \frac{1}{A+1}, \quad (S5)$$

where the aspect ratio $A = a/b$. In the case of $A = 1$, a circular biased region, everywhere on the boundar$K_U = -3/2 \, \lambda_S \, Y_{Ni} \, \gamma$ y. For $A > 1$ the maximum and minimum strain respectively are equally above and below the anisotropy value for a circular electrode. The ratio of the maximum to minimum anisotropy on the boundary is equal to the aspect ratio $A$, and the maximum anisotropy is at the most strongly curved part of the boundary.

The uniform interior magnetic anisotropy can be written in terms of the maximum and minimum anisotropy energies on the boundary. Again following Jaswon's solution, the anisotropy energy inside the biased region can be written

$$K_U^{int} = -\frac{3}{2} \lambda_{sat} E_{Ni} \big(e_x^{int} - e_y^{int}\big) = -\frac{3}{2} \lambda_{sat} E_{Ni} \, \gamma \frac{A-1}{A+1}$$

$$K_U^{int} = \frac{1}{2}\big(K_U^{max} - K_U^{min}\big).$$

At large A, $K_U^{min}$ goes to zero and the interior strain-induced anisotropy is half of that just outside of the most highly curved boundary.

Far from the boundary the magnitude of the strain decreases as $1/r^2$. To see this, let $\eta = \pi/4$ in equations (S1) and (S2) and simplify to

$$\varepsilon_{\xi\xi} = \frac{\gamma}{2}\frac{ab}{a^2-b^2}\frac{1}{\cosh 2\xi} = \frac{\gamma}{2}\frac{ab}{a^2-b^2}\frac{1}{\cosh^2\xi + \sinh^2\xi} = \frac{\gamma}{4}\frac{ab}{r^2}.$$

Although this was derived for a particular value of $\eta$ it holds for arbitrary $\eta$, as contours of constant $\xi$ are circular at large $\xi$, and strains are constant along the boundary of circular biased regions.

## Supplemental Magnetoelectric Characterization

Voltage Control of Magnetic Anisotropy Energy

Electric field induced rotation of magnetic anisotropy was measured in Figs. 2 and 3. We also measured modulation of anisotropy strength using a 300 μm by 200 μm rectangular biased region covered by continuous Ni (**Fig. S7**). This is the same type of device as was measured in Figure 2, but on a different sample in which there was Ni anisotropy in the as-grown state. Here, the as-grown magnetic anisotropy was measured to be along $\hat{x}_1$, the same direction as the strain-induced magnetic anisotropy. Assuming that the as-grown and strain induced anisotropies are parallel, then $dK_U/dV$ can be calculated by measuring $H_{\text{sat}}$ at a series of increasing voltages. The absolute size of the anisotropy in this device, up to 7 kJ/m³, is larger than those measured in Figs. 2 and 3. The anisotropy per bias voltage estimated from the slope is 0.7 kJ m⁻³ V⁻¹.

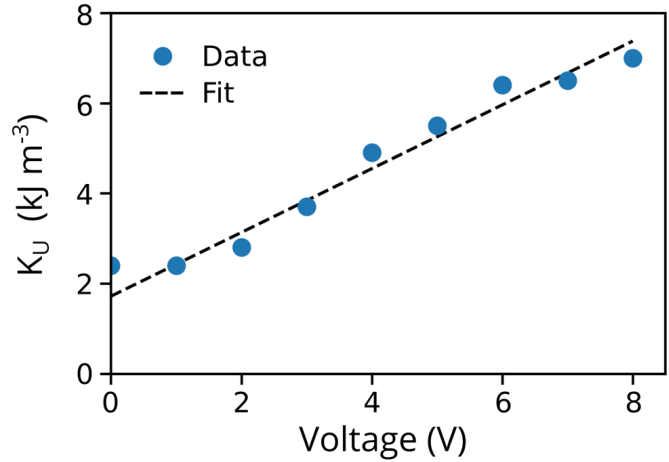

**Fig. S7.** Anisotropy energy in this device increases linearly with bias voltage across the PMN-PT, according to $H_{sat}$ estimates made from MOKE data.